\documentclass[aps,preprint]{revtex4}%
\usepackage{amsfonts}
\usepackage{amsmath}
\usepackage{amssymb}
\usepackage{graphicx}%
\begin{document}
\preprint{ }
\title[Spectral eigenvalue method]{An economical method to calculate eigenvalues of the Schr\"{o}dinger Equation. }
\author{G. Rawitscher}
\affiliation{Dept. of Physics, University of Connecticut, Storrs, CT 06268}
\author{I. Koltracht}
\affiliation{Dept. of Mathematics, University of Connecticut, Storrs, CT 06268}
\affiliation{}
\begin{abstract}
PACS number

The method is an extension to negative energies of a spectral integral
equation method to solve the Schroedinger equation, developed previously for
scattering applications. One important innovation is a re-scaling procedure in
order to compensate for the exponential behaviour of the negative energy
Green's function. Another is the need to find approximate energy eigenvalues,
to serve as starting values for a subsequent iteration procedure. In order to
illustrate the new method, the binding energy of the He-He dimer is
calculated, using the He-He TTY potential. In view of the small value of the
binding energy, the wave function has to be calculated out to a distance of
3000 a.u. Two hundred mesh points were sufficient to obtain an accuracy of
three significant figures for the binding energy, and with 320 mesh points the
accuracy increased to six significant figures. An application to a potential
with two wells, separated by a barrier, is also made.

\end{abstract}
\startpage{1}
\endpage{ }
\maketitle

\section{Introduction}

The much used differential Schr\"{o}dinger equation\ is normally solved by
means of a finite difference method, such as Numerov or Runge-Kutta, while the
equivalent integral Lippmann-Schwinger (LS) equation is rarely solved. The
reason, of course, is that the former is easier\ to implement than the
latter.\ However, a good method for solving the LS equation has recently been
developed \cite{IEM}, and applications to various atomic systems have been
presented \cite{ESRY}, \cite{E-H}. This method, denoted as S-IEM (for Spectral
Integral Equation Method), expands the unknown solution into Chebyshev
polynomials, and obtains equations for the respective coefficients. The
expansion is called "spectral", because it converges very rapidly, and hence
is economical in the number of meshpoints required in order to attain a
prescribed accuracy. A basic and simple description of the method has now been
published \cite{CISE}, and a MATLAB implementation is also included. However,
the applications described so far refer to positive energies, i.e., to
scattering situations, while an example for negative energies, i.e., bound
states, has up to now not been provided.

Since\ the solution of many quantum-mechanical problems requires the
availability of a basis of discrete negative energy eigenfunctions (or bound
states), or of positive energy Sturm-Liouville eigenfunctions, the S-IEM has
now been adapted to also obtain eigenfunctions and eigenvalues. Since there
are situations where the commonly known eigenvalue-finding methods do not work
well, we present here a short description of our method, in the hope that it
will be useful for the physics student/teacher community.

An illustration of the method for the case of the bound state of the He-He
atomic dimer is presented. This is an interesting case since the binding
energy is very small, $1.3\ mK$ or $1.1\times10^{-7}eV$\ , and the
corresponding wave function extends out to large distances, between $1000$ to
$3000$ atomic units, depending on the accuracy required. Hence a method is
desirable that maintains accuracy out to large distances, and that can find
small eigenvalues. A commonly used method to obtain eigenvalues consists in
discretizing the Schr\"{o}dinger differential operator into a matrix form, and
then numerically obtaining the eigenvalues of this matrix. This procedure
gives good accuracy for the low-lying (most bound) eigenvalues, while the
least bound eigenvalues become inaccurate. The method described here does not
suffer from this difficulty since it finds each eigenvalue of the integral
equation iteratively \ It also provides a good search procedure for finding
initial values of the eigenvalue, required to start the iteration.

\section{The formalism.}

For negative energy eigenvalues the differential equation to be solved is%

\begin{equation}
-\frac{\hbar^{2}}{2M}\frac{d^{2}\bar{\psi}}{d\bar{r}^{2}}+(\bar{V}-\bar
{E})\bar{\psi}=0 \label{schr1}%
\end{equation}
where $\bar{r}$ is the radial distance in units of length, $\bar{V}$ and
$\bar{E}$ are the potential energy and the (negative) energy in units of
energy, respectively. This is the radial equation for the partial wave of
angular momentum $0.$ For atomic physics applications this equation can be
written in the dimensionless form%
\begin{equation}
-\frac{d^{2}\psi}{dr^{2}}+(V+\kappa^{2})\psi=0 \label{schr2}%
\end{equation}
where $r=\bar{r}/a_{0}$ is the relative distance in units of Bohr, and $V$ and
$\kappa^{2}$ are given in atomic energy units. The LS eigenvalue equation that
is the equivalent to Eq. ($\ref{schr2}$), is%
\begin{equation}
\psi(r)=\int_{0}^{T}\mathcal{G}(r,r^{\prime})V(r^{\prime})\psi(r^{\prime
})dr^{\prime} \label{LS1}%
\end{equation}
where, as is well known, the Green's function $\mathcal{G}(r,r^{\prime})$ for
negative energies $-\bar{E}=(2M/\hbar^{2})\ \kappa^{2}$ is given by%
\begin{equation}
\mathcal{G}(r,r^{\prime})=-\frac{1}{\kappa}F(r_{<})G(r_{>}^{\prime})
\label{G1}%
\end{equation}
$r_{<}$ and $r_{>}$ being the lesser and larger values of $r$ and $r^{\prime}%
$, respectively, and%

\begin{equation}
F(r)=\sinh(\kappa r),\ \ \ \ \ G(r)=\exp(-\kappa r). \label{G2}%
\end{equation}

The Eq. (\ref{LS1}) is a Fredholm integral eigenvalue equation of the first
kind. Unless the wave number $\kappa$ has a correct value, the solution does
not satisfy the boundary condition that $\psi(r)$ decay exponentially at large
distances. As shown by Hartree many years ago, a method of finding a correct
value of $\kappa$ is to start with an initial guess $\kappa_{s}$ for $\kappa$,
divide the corresponding (wrong) wave function into an "out" and and "in"
part, and match the two at an intermediary point $T_{M}$. The $out$ part
$\psi_{O}$ is obtained by integrating ($\ref{LS1}$) from the origin to an
intermediate radial distance $T_{M}$, and $\psi_{I}$ is the result of
integrating ($\ref{LS1}$) from the upper limit of the radial range $T$ inward
to $T_{M}$. For the present application the integration method is based on the
S-IEM, described in Appendix 1. The function $\psi_{0}$ is renormalized so as
to be equal to $\psi_{I}$ at $r=T_{M}$ and its value at $r=T_{M}$ is denoted
as $\psi_{M}$. The derivatives with respect to $r$ at $r=T_{M}$ are
calculated, as described in Appendix 1, and are denoted as $\psi_{0}^{\prime}$
and $\psi_{I}^{\prime}$, respectively. The new value of the wave number
$\kappa_{s+1}$ is given in terms of these quantities as%
\begin{equation}
\kappa_{s+1}=\kappa_{s}-(Iter)_{s} \label{DELT1}%
\end{equation}
where
\begin{equation}
(Iter)_{s}=\frac{1}{2\kappa_{s}}\frac{\psi_{M}(\psi_{0}^{\prime}-\psi
_{I}^{\prime})_{M}}{\int_{0}^{T_{M}}\psi_{0}^{2}dr+\int_{T_{M}}^{T}\psi
_{I}^{2}dr} \label{Iter}%
\end{equation}
Equations (\ref{DELT1}) and (\ref{Iter}) can be derived by first writing
($\ref{schr2}$) for the exact wave function $\psi_{E}$ (using $\kappa_{\infty
}$ for $\kappa_{s}$ and ($\ref{schr2}$) for the approximate wave function
$\psi_{A}=(\psi_{0}$ or $\psi_{I})$, multiplying each equation by the other
wave function, integrating over $r$, and subtracting one from another. When
$\kappa_{\infty}$ is replaced by $\kappa_{s+1}$ and $\psi_{E}$ is replaced by
$\psi_{A}$ then equations ($\ref{DELT1}$) and ($\ref{Iter}$) result.

\section{The Spectral Method}

The S-IEM procedure to evaluate $\psi_{0}$ and $\psi_{I}$ is as follows. First
the whole radial interval $0\leq r\leq T$ is divided into $m$ partitions, with
the $i$-th partition defined as $t_{i-1}\leq r\leq t_{i}$, $i=1,2,\cdots m$.
For notational convenience we denote the $i$-th partition simply as $i$. In
each partition $i$ two independent functions $y_{i}(r)$ and $z_{i}(r)$ are
obtained by solving the integral equations%
\begin{equation}
y_{i}(r)=\int_{t_{i-1}}^{t_{i}}\mathcal{G}(r,r^{\prime})V(r^{\prime}%
)y_{i}(r^{\prime})dr^{\prime}+f_{i}(r) \label{y}%
\end{equation}
and
\begin{equation}
z_{i}(r)=\int_{t_{i-1}}^{t_{i}}\mathcal{G}(r,r^{\prime})V(r^{\prime}%
)z_{i}(r^{\prime})dr^{\prime}+g_{i}(r). \label{z}%
\end{equation}
Here $f_{i}$ and $g_{i}$ are scaled forms of the functions $F$ and $G$ defined
above on the interval $i$,
\begin{equation}
f_{i}(r)=\sinh(\kappa r)\times E_{i},~~~~g_{i}(r)=\exp(-\kappa r)/E_{i},
\label{f}%
\end{equation}
and the scaling factor $E_{i}$ in each partition $i$ is given by
\begin{equation}
E_{i}=\exp(-\kappa t_{i}). \label{E_i}%
\end{equation}
Such scaling factors are needed in order to prevent the unscaled functions
$\sinh(\kappa r)$ and $\exp(-\kappa r),$ and the corresponding functions
$Y_{i}$ and $Z_{i}$ to become too disparate at large distances, which in turn
would result in a loss of accuracy. Apart from these scaling operations, the
calculation of functions $y_{i}$ and $z_{i}$ by means of expansions into
Chebychev polynomials, as well as the determination of the size of the
partition $i$ in terms of the tolerance parameter $\varepsilon$ is very
similar to the calculation of the functions $Y_{i}$ and $Z_{i}$ described in
Ref. (\cite{CISE}). The number of Chebychev polynomials in each partition is
normally taken as $N=16$. The equations (\ref{y}) and (\ref{z}) are Fredholm
integral equation of the 2nd kind, and hence are much easier to solve than the
Fredholm equations of the first kind.

The global wave function $\psi$ is given in each partition by
\begin{equation}
\psi(r)=a_{i}y_{i}(r)+b_{i}z_{i}(r). \label{ABs}%
\end{equation}
In order to obtain the coefficients $a_{i}$ and $b_{i}$ for each partitions
$i$ one proceeds similarly to "Method B" described in\ Ref. (\cite{CISE}),
that relates these coefficients from one partition to those in\ a neighboring
partition. That relation is
\begin{equation}
\left[
\begin{array}
[c]{cc}%
E_{i}/E_{i+1} & 0\\
0 & E_{i+1}/E_{i}%
\end{array}
\right]  \omega_{i+1}\left[
\begin{array}
[c]{c}%
a_{i+1}\\
b_{i+1}%
\end{array}
\right]  =\gamma_{i}\left[
\begin{array}
[c]{c}%
a_{i}\\
b_{i}%
\end{array}
\right]  , \label{RECs}%
\end{equation}
where the elements of the $2\times2$ matrices $\omega$ and $\gamma$ are given
in terms of overlap integrals $\left\langle fy\right\rangle _{i},$
$\left\langle fz\right\rangle _{i},$ $\left\langle gy\right\rangle _{i},$
$\left\langle gz\right\rangle _{i},$ of the type $\left\langle fy\right\rangle
_{i}=\int_{t_{i-1}}^{t_{i}}f_{i}(r)V(r)y_{i}(r)dr$, as is described in further
detail in the Appendix 1. This relation enables one to march outward by
obtaining $a_{O,i+1}$ and $b_{O,i+1}$ in terms of $a_{O,i}$ and $b_{O,i},$ and
inward by obtaining $a_{I,i}$ and $b_{I,i}$ in terms of $a_{I,i+1}$ and
$b_{I,i+1}.$The integration outward is started at the innermost partition
$i=1$ with $a_{O,1}=1/E_{1},$ and the integration inwards is started at the
outermost partition (ending at T), for which the coefficients $a_{m}$ and
$b_{m}$ are given as $0$ and $E_{m},$ respectively. The values of the
functions $I$ and $O$ and their derivatives at the inner matching point
$T_{M}$, as well as the integrals $\int_{0}^{T_{M}}\psi_{O}^{2}dr+\int_{T_{M}%
}^{T}\psi_{I}^{2}dr,$ required for evaluating $Iter$ in Eq. (\ref{Iter}), can
be obtained in terms of the overlap integrals described above, as is described
in Appendix 1. The iteration for the final value of $\kappa$ proceeds until
the value of $Iter$ is smaller than a prescribed tolerance. The important
question of how to find an initial value $\kappa_{0}$ of $\kappa$ is described
in the next section.

\section{Search for the initial values of $\kappa$}

Since the present method does not obtain\ all the values of the energy as the
eigenvalues of one big matrix, but rather obtains iteratively one selected
eigenvalue at a time, it is necessary to have a reliable algorithm for finding
the appropriate starting values $\kappa_{0}$ for the iteration procedure.

The present search method is based on Eq. (\ref{ABs}), according to which the
solution $\psi$ in a given partition $i$ is made up of two parts, $y_{i}(r)$
and $z_{i}(r).$ In the radial regions where the potential is small compared to
the energy, i.e., in the "far" region beyond the outer turning point, the
functions $y_{i}(r)$ and $z_{i}(r)$\ are nearly equal to the driving terms $f$
and $g$ of the respective integral equations (\ref{y}) and (\ref{z}). Hence,
for negative energies, according to Eqs. (\ref{f}), \ in the "far" region
$y_{i}(r)$ has an exponentially increasing behavior, while\ $z_{i}(r)$ is
exponentially decreasing. For the correct bound state energy eigenvalue the
solution $\psi$ has to decrease exponentially at large distances, and hence
the coefficient $a_{i}$ \ in Eq. (\ref{ABs}) has to be zero for the last
partition $i=m$. Hence, as a function of $\kappa$ the coefficient $a_{m}$ goes
through zero at a value of $\kappa$ equal to one of the the bound state energies.

Based on the above considerations, the search procedure for the initial value
$\kappa_{0}$ is as follows: A convenient grid of equispaced $\kappa_{s}$
values is constructed, $s=1,2,...$ and for each $\kappa_{s}$ the integration
\ "outward" for the wave function is carried out to $T_{M}\lesssim T$ , but
$Iter$ is not calculated. The value of $T_{M}$\ is selected such that the
potential $V$\ is less than the expected binding energy. The values of the
coefficient $a_{O,i_{M}}$ for the last partition $i_{M}$ are recorded, and the
values of $\kappa_{s}$\ for which $a_{O,i_{M}}$ changes sign are the desired
starting values $\kappa_{0}$ for the iteration procedure. The numerical
example, given in the sections describing the calculation of the $He-He$ bound
state, shows that this search method is very reliable.

\subsection{The Numerical Code}

The code was written in MATLAB, and is available from the authors both in
MATLAB and in FORTRAN versions. The code that performs the iterations is
denoted as $Iter\_neg\_k,$ and the search code for finding the starting values
$\kappa_{0}$ is denoted as $Searchab\_neg\_k.$ The subroutines for both codes
are the same. The validity of the code was tested by comparing the resulting
binding energy with a non-iterative spectral algorithm that obtains the
eigenvalues of a matrix. The potential used for this comparison was an
analytical approximation to the $He-He$ potential $TTY$ \cite{TTY}, described
in the next section. The comparison algorithm expands the wave function from
$R_{START}$ to $T$ (no partitions) in terms of scaled Legendre polynomials up
to order $N$. The operator $-d^{2}/dr^{2}+V$ is discretized into a matrix at
zeros of the Legendre polynomial of order $N+1$. The boundary conditions that
the wave function vanishes at both $R_{START}$ and at $T$ are incorporated
into the matrix, and the eigenvalues of the matrix are calculated. The
agreement between the two codes for the binding energy was good to 6
significant figures.

In the test-calculation for the $He-He$ dimer binding energy described below,
the convergence rate of the iterations, the stability with respect to the
value of a repulsive core cut-off parameter, and also the number of
mesh-points required for a given input value of the tolerance parameter will
be examined. A bench-mark calculation of the dimer binding energy is also
provided for students that would like to compare their method of calculation
to ours. In these calculations the TTY potential is replaced by an analytical
approximation that is easier to implement.

\section{Application to the $He-He$ dimer}

The He-He dimer is an interesting molecule, because, being so weakly bound, it
is the largest two-atom molecule known. The He-He interaction, although weak,
does influence properties such as the superfluidity of bulk He II, of He
clusters, the formation of Bose-Einstein condensates, and the calculation of
the He trimer. In 1982 Stwalley \emph{et al} \cite{STWALLEY} were the first to
conjecture the existence of a He-He dimer. The first experimental indication
of the dimer's existence was found in 1993 \cite{LUO}, and since 1994 it was
explored by means of a series of beautiful diffraction experiments. Through
these diffraction experiments not only has the existence of the dimer, but
also that of the trimer, been unequivocally demonstrated and an indication of
the spatial extent of these molecules has also been obtained \cite{SCHOEL},
\cite{Grisenti}, \cite{BRUHL}. Various precise calculations of the He-He
interaction have subsequently been performed \cite{THEO} and the
corresponding\ theoretical binding energies of the dimer (close to
$1.3\ mK\approx1.1\times10^{-7}eV$, see Table 1 in Ref. \cite{SANDH}) and the
trimer (the ground state of the trimer is close to $126mK$, see for instance
Ref. \cite{SOFIANOS}) agree with experiment to within the experimental
uncertainty. The wave function of the He dimer or trimer extends out to large
distances (several thousand atomic units), the binding energy is very weak,
and the transition from the region of the large repulsive core to the weak
attractive potential valley is very abrupt. For these reasons the dimer (or
trimer) calculations involving He atoms require good numerical accuracy, and
therefore was chosen as a test case for our new algorithm.

The transition from Eq. (\ref{schr1}) to the dimensionless Eq. (\ref{schr2}),
is accomplished by transforming the potential and the energy into
dimensionless quantities as follows%
\begin{equation}
V=Q\bar{V} \label{v1}%
\end{equation}%
\begin{equation}
\kappa^{2}=-Q\bar{E} \label{kappa1}%
\end{equation}
where $Q$ is a normalization constant, defined in Appendix 2. For the case of
two colliding $He$ atoms interacting via the $TTY$ potential we take the mass
of the He atom as given in Ref. \cite{SANDH}, for which the value of $Q$ is
$7295.8356.$ For the calculations involving our analytical fits to the $TTY$
potential, we take for $Q$ the value $7296.3.$

\begin{figure}
[ptb]
\begin{center}
\includegraphics[
natheight=16.333700in,
natwidth=21.200001in,
height=3.1298in,
width=4.0551in
]%
{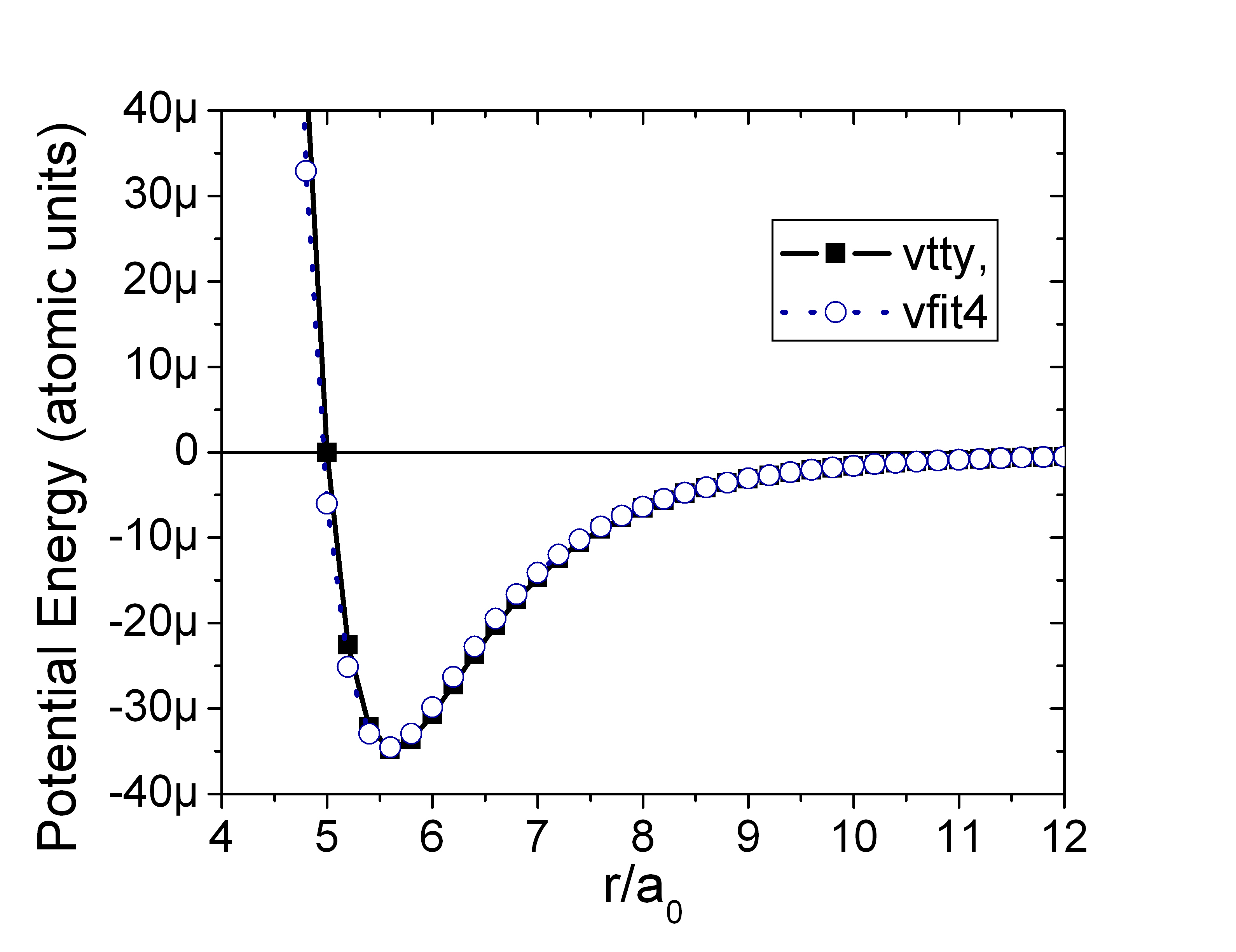}%
\caption{The "TTY" He-He potential given by Tang, Toennies, and Yiu
\cite{TTY}, and the fit FIT 4 , as a function of distance.}%
\label{FIG1}%
\end{center}
\end{figure}

The $TTY$ potential \cite{TTY}, and one analytic fit, are shown in Fig.
(\ref{FIG1}). The repulsive core goes out to about $5\ a_{0}$ and the%
%TCIMACRO{\FRAME{ftbpFU}{4.0551in}{3.1298in}{0pt}{\Qcb{The "TTY" He-He
%potential given by Tang, Toennies, and Yiu \cite{TTY}, and the fit FIT 4 , as
%a function of distance.}}{\Qlb{FIG1}}{tty2_fit4.jpg}%
%{\special{ language "Scientific Word";  type "GRAPHIC";
%maintain-aspect-ratio TRUE;  display "USEDEF";  valid_file "F";
%width 4.0551in;  height 3.1298in;  depth 0pt;  original-width 21.2in;
%original-height 16.3337in;  cropleft "0";  croptop "1";  cropright "1";
%cropbottom "0";  filename 'TTY2_fit4.JPG';file-properties "XNPEU";}}}%
%BeginExpansion
%EndExpansion
subsequent attractive valley reaches its maximum depth of $3.5\times10^{-5}au$
(approximately $10^{-3}eV$) near $r\simeq5.6\ a_{0}$. This attractive
potential valley then decays slowly over large distances approximately like
$r^{-6}$. The corresponding energy of the bound state is $\simeq-10^{-7}ev$
\cite{TTY}. In the units defined in $(\ref{schr2})$ the potential valley has a
depth of $0.26$ and the binding energy has the value of $3.04\times10^{-5}$.
The bound state wave function peaks near $r=10\ a_{0}$ and decays slowly from
there on. The outer turning point occurs near $30\ a_{0}$; the value of the
wave function at $r=2500\ a_{0}$ is $\simeq10^{-7}$, and at $3000\ a_{0}$ it
is $\simeq6\times10^{-9}$. The quantity $r\times\psi^{2}\ \ $has its maximum
beyond the turning point near $r=100\ a_{0}$, and the average radial
separation $\left\langle r\right\rangle =\int_{0}^{\infty}\psi^{2}\ r\ dr$ is
close to $\simeq97\ a_{0}.$

\subsection{Results for the $TTY$ Potential}

The $TTY$ He-He potential is calculated by means of Fortran code provided by
Franco Gianturco \cite{Franco}, and modified \emph{at hoc } for small
distances (less than $1\ a_{0}$) so that it maintains the repulsive core
nature. The potential is "cut off" at a distance $R_{cut}$ so that for $r\leq
R_{cut}$, $V(r)=V(R_{cut})$. The S-IEM calculation starts at $r=0$ and extends
to $T=3,000$ $a_{0.}$ The intermediary matching point is $T_{M}=7\ a_{0}$. The
dependence of the eigenvalue on $R_{cut},$ and the rate of convergence of the
iterations, are described in Appendix 3. Our choice for the value of
$R_{cut}=2.5\ a_{0}$, of $T=3,000$, and of the tolerance parameter
$\varepsilon=10^{-12}$ is such that the\ numerical stability of our results is
better than $12$ significant figures.

Our value for the binding energy is compared with that of other calculations
in Table \ref{TABLE2}.%
%TCIMACRO{\TeXButton{B}{\begin{table}[tbp] \centering}}%
%BeginExpansion
\begin{table}[tbp] \centering
%EndExpansion%
\begin{tabular}
[c]{|l|l|l|}\hline
& $B.E.(mK)$ & $<r>(nm)$\\\hline
Present & $\mathbf{1.31461}$ & $\mathbf{5.1607}$\\\hline
Ref. \cite{SANDH} & $\mathbf{1.30962}$ & \\\hline
Ref. \cite{TTY} & $\mathbf{1.316}$ & \\\hline
Experiment \cite{Grisenti} & $\mathbf{0.9-1.4}$ & $\mathbf{5.2\pm0.4}$\\\hline
\end{tabular}
\caption{Comparison of the He-He Binding Energies obtained by various authors.}\label{TABLE2}%
%TCIMACRO{\TeXButton{E}{\end{table}}}%
%BeginExpansion
\end{table}%
%EndExpansion
. The comparison shows good agreement of our result with the literature. The
difference between our S-IEM result and that of Ref. \cite{SANDH} could well
be due to a slightly different choice of the parameters that determine $TTY.$

\subsection{Numerical Properties of the S-IEM.}

In order to examine the nature of the partition distribution and the resulting
accuracy as a function of the tolerance parameter $\varepsilon$ and also in
order to provide a bench-mark calculation, the $TTY$ potential was replaced by
an analytical approximation defined in the equation below.%
\begin{align}
V(r) &  =p_{1}\exp(-(r-p_{2})/p_{3})\times\left[  2-\exp(-(r-p_{2}%
)/p_{3})\right]  \nonumber\\
&  -p_{6}\times(r^{-5.807})/\left\{  1+\exp\left[  -p_{5}(r-p_{4})\right]
\right\}  .\label{FIT_EQ}%
\end{align}
The parameters $p_{1}$ to $p_{6}$ for two fits, denoted FIT 3 and FIT 4,\ are
given in Table 2. The resulting potential is in atomic units; $r,p_{2},p_{3}$
and $p_{4}$ are in units of $a_{0}$; $p_{5}$ is in units of $a_{0}^{-1}$ and
$p_{1}$ and $p_{6}$ are in atomic energy units. For all calculations involving
these analytical fits, $Q$ defined in Eq. (\ref{v1}), has the value $7296.3.$
%TCIMACRO{\TeXButton{B}{\begin{table}[tbp] \centering}}%
%BeginExpansion
\begin{table}[tbp] \centering
%EndExpansion%
\begin{tabular}
[c]{|c|c|c|}\hline
parameters & FIT 3 & FIT 4\\\hline
$p_{1}$ & -3.4401e-5 & -2.930 e-5\\\hline
$p_{2}$ & 5.606 & 5.590\\\hline
$p_{3}$ & 0.8695 & 0.8511\\\hline
$p_{4}$ & 7.657 & 7.5892\\\hline
$p_{5}$ & 1.750 & 0.95608\\\hline
$p_{6}$ & 0.6784 & 0.89098\\\hline
\end{tabular}
\caption{Parameters for two analytic fits to the TTY potential }\label{TABLE4}%
%TCIMACRO{\TeXButton{E}{\end{table}} }%
%BeginExpansion
\end{table}
%EndExpansion
Fits 3 (4) produce a more (less) repulsive core than TTY, and is more (less)
attractive in the region of the potential minimum.

Our algorithm automatically chooses the size of the partitions such that the
error in the functions calculated in each partition does not exceed the
tolerance parameter $\varepsilon.$ At small distances the density of
partitions is very high, but beyond $500\ a_{0}$ the size of the partitions
increase to about $440\ a_{0}$. In the region near the repulsive core the
partitions are\ approximately $0.5$ $a_{0}$ wide, but there is a region in the
vicinity of $R_{cut\text{ }}$ where they crowd together much more. The latter
is illustrated in Fig. (\ref{FIG9}) for FIT 4, for $R_{cut}=2.5\ a_{0},$ for
various values of the tolerance parameter $\varepsilon$. In Table \ref{TABLE6}
the corresponding accuracy of the binding energy is displayed, for the case
that the $He-He$ potential energy is given by Fit 4. It is noteworthy that the
number of reliable significant figures in $\kappa$ tracks faithfully the value
of the tolerance parameter, as is shown in Table \ref{TABLE6}.%
%TCIMACRO{\FRAME{ftbpFU}{2.9879in}{2.3047in}{0pt}{\Qcb{Partition distribution
%in the radial region up to 4 $a_{0}$ for three different values of the
%tolerance parameter. The value of the latter is listed in the legend and in
%Table \ref{TABLE6}. The potential is given by FIT 4, described in the text and
%the value of $R_{cut}=2.5\ a_{0}.$ The total number of partitions for each
%case is given by the numbers near the curves.}}{\Qlb{FIG9}}{spart(tol).jpg}%
%{\special{ language "Scientific Word";  type "GRAPHIC";
%maintain-aspect-ratio TRUE;  display "USEDEF";  valid_file "F";
%width 2.9879in;  height 2.3047in;  depth 0pt;  original-width 21.2in;
%original-height 16.3337in;  cropleft "0";  croptop "1";  cropright "1";
%cropbottom "0";  filename 'Spart(tol).JPG';file-properties "XNPEU";}}}%
%BeginExpansion
\begin{figure}
[ptb]
\begin{center}
\includegraphics[
natheight=16.333700in,
natwidth=21.200001in,
height=2.3047in,
width=2.9879in
]%
{Spart(tol).jpg}%
\caption{Partition distribution in the radial region up to 4 $a_{0}$ for three
different values of the tolerance parameter. The value of the latter is listed
in the legend and in Table \ref{TABLE6}. The potential is given by FIT 4,
described in the text and the value of $R_{cut}=2.5\ a_{0}.$ The total number
of partitions for each case is given by the numbers near the curves.}%
\label{FIG9}%
\end{center}
\end{figure}
%EndExpansion%
%TCIMACRO{\TeXButton{B}{\begin{table}[tbp] \centering}}%
%BeginExpansion
\begin{table}[tbp] \centering
%EndExpansion%
\begin{tabular}
[c]{|l|l|l|l|}\hline
$Tol$ & $\kappa\times10^{3}~(a_{0})^{-1}$ & $M$ & $No.\ of\ Meshpts$%
\\\hline\hline
$10^{-12}$ & $5.0817542$ & $47$ & $652$\\\hline
$10^{-6}$ & $5.0817461$ & $19$ & $275$\\\hline
$10^{-3}$ & $5.0776$ & $13$ & $208$\\\hline
\end{tabular}
\caption{Accuracy of the  wave number (it is the square root of the binding energy) as a function of the tolerance parameter. The total number of partitions for each case is denoted by M. The corresponding partition distributions are displayed in Fig. 4}\label{TABLE6}%
%TCIMACRO{\TeXButton{E}{\end{table}}}%
%BeginExpansion
\end{table}%
%EndExpansion
\medskip\ 

\subsection{The Search for the Starting Values $\kappa_{0}.$}

An example of the search procedure is given in Table \ref{TABLE7}, for a
potential given by Fit 4, Eq. (\ref{FIT_EQ}), multiplied by the factor
$\lambda=$ $20$. The mesh of $\kappa$ values starts at $\kappa=-2\ $and
proceeds by steps of $\Delta\kappa=0.05$ until $\kappa=-0.05$ (all in units of
$a_{0}^{-1}$). The mesh values of $\kappa$ for which the coefficient $a_{m}$
of $y_{m}(r)$ changes sign are shown in the first column of Table
\ref{TABLE7}, and the corresponding iterated value of $\kappa$ is shown in the
third column. The value of $T_{M}=80\ a.u.$, and $iter=10^{-6}$. The MATLAB
computing time required for carrying out the $40$ mesh search calculations is
$3.8\ s$ on a $2$ GHz PC; the approximately $7$ iterations required for
obtaining the more precise values of each $\kappa$ shown in the third column
take approximately $1\ s.$ of computer time.%
%TCIMACRO{\TeXButton{B}{\begin{table}[tbp] \centering}}%
%BeginExpansion
\begin{table}[tbp] \centering
%EndExpansion%
\begin{tabular}
[c]{|l|l|l|l|}\hline
$\kappa-Mesh$ & sign of $a_{m}$ & $\kappa-Iterated$ & \# of
nodes\\\hline\hline
$1.70$ & $+\rightarrow-$ & $1.7028$ & $0$\\\hline
$0.70$ & $-\rightarrow+$ & $0.7273$ & $1$\\\hline
$0.05$ & $+\rightarrow-$ & $0.0561$ & $2$\\\hline
\end{tabular}
\caption{Search of the wave number eigenvalues for a He-He Fit 4 potential multiplied by 20 }\label{TABLE7}%
%TCIMACRO{\TeXButton{E}{\end{table}}}%
%BeginExpansion
\end{table}%
%EndExpansion
.\newline By repeating the same procedure for different values of $\lambda$,
one can trace the $\kappa$ eigenvalues down to $\lambda=1.$
%TCIMACRO{\FRAME{ftbpFU}{3.1661in}{2.4431in}{0pt}{\Qcb{The eigenvalues of
%$\kappa$ as a function of the strength parameter $\lambda$ of the Fit 4
%Potential.}}{\Qlb{FIG8}}{lanodes.jpg}{\special{ language "Scientific Word";
%type "GRAPHIC";  maintain-aspect-ratio TRUE;  display "USEDEF";
%valid_file "F";  width 3.1661in;  height 2.4431in;  depth 0pt;
%original-width 21.2in;  original-height 16.3337in;  cropleft "0";
%croptop "1";  cropright "1";  cropbottom "0";
%filename 'LaNodes.JPG';file-properties "XNPEU";}}}%
%BeginExpansion
\begin{figure}
[ptb]
\begin{center}
\includegraphics[
natheight=16.333700in,
natwidth=21.200001in,
height=2.4431in,
width=3.1661in
]%
{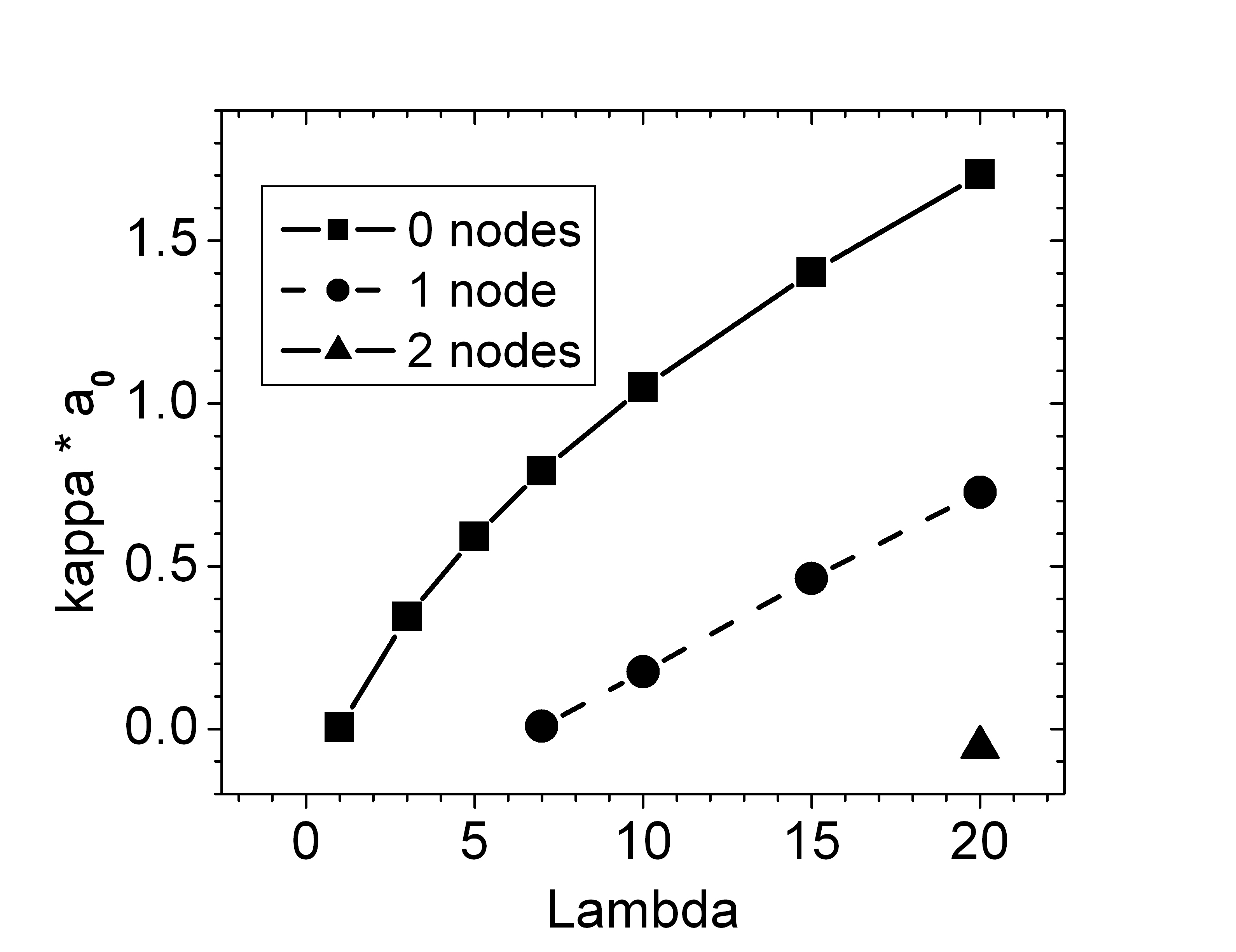}%
\caption{The eigenvalues of $\kappa$ as a function of the strength parameter
$\lambda$ of the Fit 4 Potential.}%
\label{FIG8}%
\end{center}
\end{figure}
%EndExpansion
The result, displayed in Fig. (\ref{FIG8}), shows that the values of $\kappa$
depend nearly linearly on the value of $\lambda.$ Futher searches with values
of $\lambda$ slightly less than unity showed that the code was able to find an
energy that is approximately 25 times less bound than the result for the $TTY$ potential.

In order to provide a benchmark calculation, the values of $\kappa$ obtained
with the potential of FIT 4 are listed in Table \ref{TABLE5}. The value of $Q$
is $7296.3$, the values of $T_{M}$ and $T$ are $7$ and $3,000\ a_{0},$
respectively, and $\lambda=1.0$\newline%
%TCIMACRO{\TeXButton{B}{\begin{table}[tbp] \centering}}%
%BeginExpansion
\begin{table}[tbp] \centering
%EndExpansion%
\begin{tabular}
[c]{|c|c|}\hline
$R_{cut}~(a_{0})$ & $\kappa~(a_{0})^{-1}$\\\hline
2.5 & 5.08175419E-3\\\hline
3.0 & 5.08176556E-3\\\hline
3.5 & 5.10608688E-3\\\hline
\end{tabular}
\caption{The values of the bound state wave number  for potential of FIT 4, for various values of the cut-off radius . Additional information is given in the text}\label{TABLE5}%
%TCIMACRO{\TeXButton{E}{\end{table}}}%
%BeginExpansion
\end{table}%
%EndExpansion

\section{Application to a double well potential.}

The case for which the potential has two (or more) wells separated by one (or
more) barriers offers another test for the reliability and accuracy of a
numerical procedure for obtaining eigenvalues of the Schr\"{o}dinger equation.
The reason is that the energy eigenvalues are split by a small amount,
corresponding to the situation in which the wave function located in one of
the wells has either the same or the opposite sign of the wave function
located in the adjoining well. The larger the barrier, the smaller is the
difference $\Delta E$ between the two energies, and the larger are the demands
on the numerical procedure. An interesting relaxation method for finding
energy eigenvalues contained in a prescribed interval has been described in
Ref. \cite{Carlo}. The double well potential, in the units of Eq.
(\ref{schr2}) is
\begin{equation}
V=-\Lambda x^{2}+x^{4},\text{~~}-T_{m}\leq x\leq T_{m}. \label{VDW}%
\end{equation}
The value of $\Delta E$ for the difference between the two lowest eigenvalues
were calculated here by using the S-IEM method described in this paper,
denoted as $(\Delta E)_{IEM}$and also by a matrix eigenvalue method, denoted
as $(\Delta E)_{L}.$ This method discretizes the Schr\"{o}dinger operator on
the left hand side of Eq. (\ref{schr3})
\begin{equation}
(-\frac{d^{2}}{dr^{2}}+V)\psi=E\ \psi\label{schr3}%
\end{equation}
at the zeros of a Legendre Polynomial of order $n_{L},$ and then finds all the
eigenvalues of the corresponding matrix using the standard QR algorithm. The
comparison of the results for the three largest values of $\Lambda$ is shown
in Table \ref{TABLE8}, where $(\Delta E)_{rel}$ denotes the result obtained in
Ref. \cite{Carlo}
%TCIMACRO{\TeXButton{B}{\begin{table}[tbp] \centering}}%
%BeginExpansion
\begin{table}[tbp] \centering
%EndExpansion%
\begin{tabular}
[c]{|l|l|l|l|}\hline
$\Lambda$ & $(\Delta E)_{IEM}$ & $(\Delta E)_{L}$ & $(\Delta E)_{rel}$\\\hline
$10$ & 3.02E-5 & 2.98185E-5 & 2.9821E-5\\
$12$ & 3.53E-7 & 3.508E-7 & 3.5093E-7\\
$15$ & 2E-10 & 2E-10 & 1.9499E-10\\\hline
\end{tabular}
\caption{Comparison between three different methods of calculating energy eigenvalues. The table shows the difference between the two lowest eigenvalues of the double well potential defined in this section}\label{TABLE8}%
%TCIMACRO{\TeXButton{E}{\end{table}}}%
%BeginExpansion
\end{table}%
%EndExpansion
For the S-IEM the value of the tolerance parameter was $\varepsilon=10^{-12},$
and the corresponding accuracy was sufficient to obtain the results shown in
the Table \ref{TABLE8}. However,\ it can be seen that the relaxation method is
more accurate than the S-IEM method. The difference between the results in
Table \ref{TABLE8} could well be due to differences in the choice of the value
of $T_{M}$. For the $(\Delta E)_{L}$ result, the value of $T_{M}$ was varied
between $6$ and $9$ units of length, and the number of Legendre polynomials
$n_{L}$ was varied between $200$ and $700.$ The numerical stability of the QR
algorithm is well documented in the numerical linear algebra literature. The
convergence of the Legendre discretization of the Schr\"{o}dinger operator
using finite series expansions in orthogonal polynomials, such as Legendre,
Chebyshev and others, is also well understood, as discussed for example in Ref
\cite{IEM}.

\section{Summary and Conclusions.}

An integral equation method (S-IEM) \cite{IEM} for solving the Schr\"{o}dinger
equation for positive energies has been extended to negative bound-state
energy eigenvalues. Our new algorithm is in principle very similar to an
iterative method given by Hartree in 1930, in that it guesses a binding
energy, integrates the Schr\"{o}dinger Equation inwards to an intermediary
matching point starting at a large distance, integrates it outwards from a
small distance to the same matching point, and from the difference between the
logarithmic derivatives at this point an improved value of the energy is
found. Our main innovation to this scheme is to replace the usual finite
difference method of solving the Schr\"{o}dinger equation by a method which
solves the corresponding integral (Lippman-Schwinger) equation. That method
expands the wave function in each radial partition in terms of Chebyshev
polynomials, and solves matrix equations for the coefficients of the
expansion. Increased accuracy is obtained by this procedure for three reasons:
a) the solution of an integral equation is inherently more accurate than the
solution of a differential equation; b) by using integral equations, the
derivatives of the wave function required at the internal matching points can
be expressed in terms of integrals that are more accurate than calculating the
derivatives by a numerical three- or five point formula, and c) because of the
spectral nature of the expansion of the wave function in each partition, the
length of each partition can be automatically adjusted in order to maintain a
prescribed accuracy. This last property enables the S-IEM to treat accurately
the abrupt transition of the wave function from the repulsive core region into
the attractive valley region. This feature, once applied to the solution of
the three-body problem, is also of importance in the exploration of the Efimov
states \cite{INCAO}.

To illustrate this method, the binding energy of the He dimer has been
calculated, based on the $TTY$ potential given by Tang, Toennies, and Yiu
\cite{TTY}. The result is close to the ones quoted in the literature, as
displayed in Table \ref{TABLE3}. Additional numerical properties of the S-IEM
have been explored by means of the $He-He$ example. The accuracy of the
binding energy was found to faithfully track the input value of the tolerance
parameter, as is shown in Table \ref{TABLE6}. The meshpoint economy of the
method is very good. For an accuracy of three significant figures, the number
of meshpoints needed in the radial interval between $0$ and $3,000\ a.u.$
required only $208$ mesh points. After an addition of $70$ meshpoints, the
accuracy increased to six significant figures.

One of the authors (GR) acknowledges useful conversations with F. A.
Gianturco, W. Gl\"{o}ckle,\ I. Simbotin, W. C. Stwalley, and K. T. Tang

\bigskip

{\LARGE Appendix 1:} {\LARGE Recursion Relations for the coefficients
}${\LARGE a}${\LARGE  and  }${\LARGE b}${\LARGE .}

The recursion relation between coefficients $a$ \ and $b$ , from one partition
to a neighbouring partition is given by Eq. (\ref{RECs}) in the text. The
corresponding matrices $\omega_{i}$ and $\gamma_{i}$ are given by%
\begin{equation}
\omega_{i}=\left[
\begin{array}
[c]{cc}%
0 & 1\\
1-\left\langle gy\right\rangle _{i} & -\left\langle gz\right\rangle _{i}%
\end{array}
\right]  \label{omegai}%
\end{equation}%
\begin{equation}
\gamma_{i}=\left[
\begin{array}
[c]{cc}%
-\left\langle fy\right\rangle _{i} & 1-\left\langle fz\right\rangle _{i}\\
1 & 0
\end{array}
\right]  \label{gammai}%
\end{equation}
where
\begin{equation}
\left\langle fy\right\rangle _{i}=\int_{t_{i-1}}^{t_{i}}f_{i}(r)V(r)y_{i}(r)dr
\label{ovfy}%
\end{equation}%
\begin{equation}
\left\langle fz\right\rangle _{i}=\int_{t_{i-1}}^{t_{i}}f_{i}(r)V(r)z_{i}(r)dr
\label{ovfz}%
\end{equation}%
\begin{equation}
\left\langle gy\right\rangle _{i}=\int_{t_{i-1}}^{t_{i}}g_{i}(r)V(r)y_{i}(r)dr
\label{ovgy}%
\end{equation}%
\begin{equation}
\left\langle gz\right\rangle _{i}=\int_{t_{i-1}}^{t_{i}}g_{i}(r)V(r)z_{i}%
(r)dr. \label{ovgz}%
\end{equation}

Equation (\ref{RECs}) enables one to march outward%

\begin{equation}
\left[
\begin{array}
[c]{c}%
a_{0,i+1}\\
b_{0,i+1}%
\end{array}
\right]  =(\omega_{i+1})^{-1}\left[
\begin{array}
[c]{cc}%
E_{i+1}/E_{i} & 0\\
0 & E_{i}/E_{i+1}%
\end{array}
\right]  \gamma_{i}\left[
\begin{array}
[c]{c}%
a_{0,i}\\
b_{0,i}%
\end{array}
\right]  , \label{REC0}%
\end{equation}
or inward
\begin{equation}
\left[
\begin{array}
[c]{c}%
a_{I,i}\\
b_{I,i}%
\end{array}
\right]  =\gamma_{i}^{-1}\left[
\begin{array}
[c]{cc}%
E_{i}/E_{i+1} & 0\\
0 & E_{i+1}/E_{i}%
\end{array}
\right]  \omega_{i+1}\left[
\begin{array}
[c]{c}%
a_{I,i+1}\\
b_{I,i+1}%
\end{array}
\right]  . \label{RECI}%
\end{equation}

The integration outward is started at the innermost partition $i=1$ with
\begin{equation}
\left[
\begin{array}
[c]{c}%
a_{O,1}\\
b_{O,1}%
\end{array}
\right]  =\left[
\begin{array}
[c]{c}%
1/E_{1}\\
0
\end{array}
\right]  , \label{start0}%
\end{equation}
and the integration inwards is started at the outermost partition (ending at
T), for which the coefficients $a_{m}$ and $b_{m}$ are given as
\begin{equation}
\left[
\begin{array}
[c]{c}%
a_{I,m}\\
b_{I,m}%
\end{array}
\right]  =\left[
\begin{array}
[c]{c}%
0\\
E_{m}%
\end{array}
\right]  . \label{startI}%
\end{equation}
If the calculation of positive energy Sturm-Liouville functions is envisaged,
whose asymptotic behavior is $\exp(ikr)$ and approach $0$ for $r\rightarrow0,$
then $a_{I,m}=i$ and $b_{I,m}=1,$ while $a_{O,1}=1$ and $b_{O,1}=0$

The values of the functions $y$ and $z$ and their derivatives at upper and
lower end-points $t_{i}$ and $t_{i-1}$ of partition $i,$ required in the
evaluation of Eq. (\ref{Iter}), are obtained from integral equations that
these functions obey. The result is \cite{CISE}
\begin{equation}
y_{i}(t_{i})=f_{i}(t_{i})-\left\langle fy\right\rangle _{i}g_{i}(t_{i}),
\label{yti}%
\end{equation}%
\begin{equation}
z_{i}(t_{i})=g_{i}(t_{i})(1-\left\langle fz\right\rangle _{i}), \label{zti}%
\end{equation}%
\begin{equation}
y_{i}(t_{i-1})=f_{i}(t_{i-1})(1-\left\langle gy\right\rangle _{i}),
\label{ytim1}%
\end{equation}%
\begin{equation}
z_{i}(t_{i-1})=g_{i}(t_{i-1})-f_{i}(t_{i-1})\left\langle gz\right\rangle _{i}.
\label{ztim1}%
\end{equation}
Expressions for the derivatives of $y$ and $z$ at upper and lower end-points
$t_{i}$ and $t_{i-1}$ of partition $i$ are obtained by replacing functions $f$
and $g$ by their respective derivatives in the above equations. Since
derivatives of the functions $f$ and $g$ are given analytically, the values of
the derivatives of $y$ and $z$ at the end-points are obtained without loss of
accuracy, contrary to what is the case when finite difference methods are
employed\bigskip

{\LARGE Appendix 2: Units}

The transition from Eq. (\ref{schr1}) to the dimensionless Eq. (\ref{schr2})
is accomplished by transforming the potential and the energy into
dimensionless quantities according to Eqs. (\ref{v1}) and (\ref{kappa1}). The
normalization constant is given by%
\begin{equation}
Q=\frac{2M}{\hbar^{2}}a_{0}^{2}\times2\mathbb{R}=\frac{2M}{m_{e}}, \label{Qt}%
\end{equation}
where $a_{0}$ is the Bohr radius, $2\mathbb{R}$ is the atomic energy unit
($\mathbb{R\simeq}13.606eV$), $\hbar$ is Plank's constant divided by $2\pi$,
$M$ is the reduced mass of the colliding atoms , and $m_{e}$ is the mass of
the electron.

For the case of two colliding He atoms interacting via the $TTY$ potential we
take the mass of the $He$ atom as given in Ref. \cite{SANDH}, i.e., $\hbar
^{2}/M_{^{4}He}=12.12\ K\ \mathring{A}^{2}$ for which the value of $Q$ is%

\begin{equation}
Q=7295.8356 \label{QN}%
\end{equation}
Once $\kappa^{2}$ is obtained as the eigenvalue of equation ($\ref{schr2}$),
then the corresponding value of $\bar{E}$ in units of $eV$ is given by%

\begin{equation}
\bar{E}= - \frac{\kappa^{2}}{Q} \times(27.211396) \ \ eV \label{EeV}%
\end{equation}
It is also useful to express the energy in units of the Boltzman constant,
denoted by $K$ in atomic language. In this case $\bar{E}$ is given as%

\begin{equation}
\bar{E}=-\frac{\kappa^{2}}{Q}\times\frac{27.211396}{8.617385\times10^{-5}%
}\ \ K. \label{EK}%
\end{equation}

\bigskip

{\LARGE Appendix 3: Accuracy Considerations}

The quantities required for Eqs. (\ref{REC0}) and (\ref{RECI}) are known to
the same accuracy as the functions $y$ and $z$ in each partitions, given by
the value of the tolerance parameter $\varepsilon$. The propagation of the
coefficients $a_{i}$ and $b_{i}$ across the partitions involves as many matrix
inversions and multiplications in Eqs. (\ref{REC0}) and \ref{RECI}) as there
are partitions, and thus the accuracy of $\kappa_{s}$ \ for each iteration,
given by Eq. (\ref{Iter}), is reduced by $tol\times$ number of partitions. The
number of partitions is approximately $30$,\ hence for $\varepsilon=10^{-12}$
the accuracy of the final wave number eigenvalue $\kappa$ is expected to be
better than $10^{-10}$.

The rate of convergence of the iterations is shown in Table \ref{TABLE3}.
%TCIMACRO{\TeXButton{B}{\begin{table}[tbp] \centering}}%
%BeginExpansion
\begin{table}[tbp] \centering
%EndExpansion%
\begin{tabular}
[c]{|c|c|c|}\hline
$s$ & $\kappa_{s}$\ \ ($a_{0}$)$^{-1}$ & $Iter_{s}$ (from (\ref{Iter}%
))\\\hline
$0$ & $\mathbf{3}.0\ E-3$ & $-2.5002592843\ E-3$\\\hline
$1$ & $\mathbf{5.5}002592823\ E-3$ & $-1.0967998971\ E-5$\\\hline
$2$ & $\mathbf{5.511227}2813\ E-3$ & $-2.0105203008\ E-10$\\\hline
$3$ & $\mathbf{5.5112274823\ }E-3$ & $-4.9700035857\ E-16$\\\hline
\end{tabular}
\caption{Convergence of the iterations for the wave number . The quantitie after the letter E denote the powers of 10 by which the quantities are to be multiplied. }\label{TABLE3}%
%TCIMACRO{\TeXButton{E}{\end{table}}}%
%BeginExpansion
\end{table}%
%EndExpansion
. The sensitivity of the binding energy to the values of $R_{cut}$ is given in
Table \ref{TABLE1}.%
%TCIMACRO{\TeXButton{B}{\begin{table}[tbp] \centering}}%
%BeginExpansion
\begin{table}[tbp] \centering
%EndExpansion%
\begin{tabular}
[c]{|l|l|l|}\hline
$R_{cut}(a_{0})$ & $B.E.(m.K)$ & $<r>(a_{0})$\\\hline
$2.0$ & $\mathbf{1.3146101}$ & $\mathbf{97.7419}$\\\hline
$2.5$ & $\mathbf{1.3146101}$ & $\mathbf{97.7419}$\\\hline
$3.0$ & $\mathbf{1.3146143}$ & $\mathbf{97.7418}$\\\hline
$3.5$ & $\mathbf{1.3219315}$ & $\mathbf{97.4935}$\\\hline
\end{tabular}
\caption{Sensitivity of the He-He Binding Energy to the value of the cuting-off radius $R_{cut}$}\label{TABLE1}%
%TCIMACRO{\TeXButton{E}{\end{table}}}%
%BeginExpansion
\end{table}%
%EndExpansion
. \ The table shows that the repulsive core has a non-negligible effect in the
$7th$ significant figure beyond $2.5\ a_{0}$.

\end{document}